*applied system innovation*

MDPI

*Article*

# Effect of Surface Roughness on Early Stage Oxidation Behavior of Ni-Base Superalloy IN 625

Wojciech J. Nowak

Department of Materials Science, Faculty of Mechanical Engineering and Aeronautics, Rzeszow University of Technology, al. Powstancow Warszawy 12, 35-959 Rzeszow, Poland; w.nowak@prz.edu.pl; Tel.: +48-17-743-2375

**Abstract:** In the present work the effect of surface roughness on oxidation behavior during the early stages of high temperature exposure of Ni-base superalloy IN 625 is described. The surface roughness was described using standard contact profilometer as well as novel method, fractal analysis. It was found that the different surface preparation resulted in a difference in roughness with a parameter increase of at least one order of magnitude for the ground sample as compared with the polished sample. The oxidation test was performed in a horizontal tube furnace. Post-exposure analyses including glow discharge optical emission spectrometry (GD-OES) and scanning electron microscopy (SEM), which revealed that grinding lowers the oxidation kinetics of IN 625 from $1.76 \times 10^{-12}$ cm$^2 \cdot$s$^{-1}$, obtained for polished sample, to $9.04 \times 10^{-13}$ cm$^2 \cdot$s$^{-1}$. It was found that surface preparation influences the oxide scale composition and morphology. The hypothesis explaining the mechanism responsible for the changes in oxidation behavior is proposed as well.

**Keywords:** Ni-base superalloy; IN 625; oxidation kinetics; surface roughness; fractal analysis

## 1. Introduction

The materials used in the hottest parts of stationary gas turbines or jet engines face very aggressive atmospheres at high temperature. Therefore, these materials should fulfill a number of requirements, such as, for example, high creep strength and high oxidation resistance at a wide range of operating temperatures, environments, and loading conditions. Ni-base superalloys exhibit excellent mechanical properties at high temperature, due to their microstructure. They possess also good oxidation resistance due to the addition of elements forming a protective oxide layer, like Al or Cr [1]. IN 625 is one of the Ni-base superalloys which is widely used for application at high temperatures. According to the oxidation map developed by Giggins and Pettit [2], IN 625 can be classified as a chromia forming alloy, as it forms a slow growing $Cr_2O_3$ scale. For better protection against oxidation, Ni-base superalloys can be covered by additional oxidation resistant layers such as MCrAlY type coatings [3] or β-NiAl [4], which at high temperatures form the most protective alumina oxide scale. However, manufacturing of those coatings generate additional costs.

Literature analysis shows that the oxidation behavior of high temperature material can be controlled by the surface conditions. Giggins and Pettit investigated the influence of surface deformation and/or grain size of Ni-Cr alloy [5]. They found that the grit blasted material showed similar oxidation behavior to that with fine grained material. The authors claimed that the grit blasted material has recrystallized and the grain boundaries enhanced Cr diffusion along them. Leistikow et al. investigated the influence of grain boundary density introduced by the cold working of Alloy 800 [6]. The authors found that the higher grain boundary density leads to smaller mass gain, while with decreasing grain boundaries density, the obtained mass gain increased. Sudbrack et al. investigated LDS-1101_HF and CMSX-4 with different surface conditions [7]. They observed strong a correlation between the surface preparation and type of formed oxide scale, namely, on ground surface, thin $Al_2O_3$ scale was observed, while on polished sample, a much thicker NiO outer scale with a subscale of $Cr_2O_3$-spinel-$Al_2O_3$ was found. Nowak et al. investigated the role of surface finishing on oxidation kinetics of IN 713 C [8]. It was found that chromium rich oxide formed on polished surfaces, while alumina oxide formed on ground surfaces.

The literature study revealed that there are limited works describing the influence of surface condition on the oxidation behavior of IN 625. Garcia-Fresnillo et al. [9] studied oxidation behavior of IN 625 during the long term oxidation in steam at temperature range 700–800 °C. The samples of IN 625 were prepared by grinding up to 1200 grit SiC paper. Staszewska et al. [10] compared the oxidation mechanism and kinetics of IN 617 and IN 625 during exposure at 1200 °C, with the sample prepared by grinding up to 2500 grit SiC paper. Kumar et al. [11] studied the oxidation behavior of IN 625 polished up to mirror finishing at temperature range 500–1250 °C. Vesel et al. [12] studied oxidation behavior of IN 625 upon treatment with oxygen and hydrogen plasma at temperatures up to 1330 °C. The surface of the alloy was prepared by grinding using fine-sandpaper grade. However, no single work dealing with the influence of the surface preparation on the oxidation behavior of IN 625 has been found.

Therefore, the aim of the present work is to investigate the influence of surface preparation, namely polishing up to 1 μm and grinding up to 220 grit SiC paper, on oxidation kinetics and oxide scale morphology formed on IN 625 during early stages (up to 24 h) of exposure at 950 °C in air. The oxidation behavior is correlated with the surface roughness. A mechanism of oxide formation on polished and ground surfaces is proposed as well.

**2. Materials and Methods**

In the present work a commercially available Ni-base superalloy IN 625, with nominal composition given in Table 1, was investigated.

**Table 1.** Nominal chemical composition of IN 625 expressed in wt %.

| Elements (wt %) | | | | | | | | | |
|---|---|---|---|---|---|---|---|---|---|
| Ni | Cr | Co | Mo | Nb | Al | Ti | Fe | Mn | C |
| BASE | 21.5 | 1 | 9 | 3.65 | 0.4 | 0.4 | 5 | 0.5 | 0.1 |

From the rods of IN 625, 2 mm thick coupons with diameter of 20 mm were machined. The surfaces of the samples were prepared with two methods, namely grinding up to 220 grit SiC paper and polishing using a 1 μm diamond suspension. To control the reproducibility of the test results, two samples per each surface condition were prepared. After surface preparation, samples were ultrasonically cleaned in ethanol. The roughness of the samples was measured using two methods: by traditional roughness measurement using contact profilometer HOMMEL Werke T8000 and by fractal analysis of the cross-sections produced from the ground and polished sample cross-section images. The surface roughness of each sample was evaluated by commonly used parameters such as arithmetic average height ($R_a$), average distance between the highest peak and lowest valley ($R_z$), and maximum height of the profile ($R_{max}$), of which the calculation procedures are described in [13,14]. The analysis by contact profilometer was done under the following parameters: the traverse length ($L_t$) was 4 mm, linear speed of stylus ($V_t$) was 0.5 mm/s, and cut off length was 0.8 mm. Fractal analysis of polished and ground samples was performed on the cross-sections of as-prepared alloys in accordance with the procedure given by Nowak et al. [15] using Sfrax 1.0 software [16]. The procedure of roughness analysis using fractal analysis included the following:

(a) Image acquisition by optical microscope,
(b) Converting images into binary scale images (alloy—white, resin—black),
(c) Digitalization of binary scale images into 2-D x-y data sets with constant x step,
(d) performing fractal analysis using Sfrax 1.0 software [16].

As described in [10] based on fractal analysis the plots showing the relative length as function of scale length can be drawn. Also the following characteristic parameters describing the measured surface roughness profiles can be derived [17,18]:

- D (Fractal Dimension), given by the Equation (1):

$$D = 1 + |slope|, \qquad (1)$$

- LSFC (Length-Scale Fractal Complexity), given by the Equation (2):

$$LSFC = 1000\,(D-1), \qquad (2)$$

- Relative length at a given scale ($L_R$)
- Smooth-rough crossover (SRC)

Rougher surfaces possess higher values of D, LSFC, and relative length at a given scale, while the value of SRC indicates the scale at which roughness starts to prevail (e.g., on μm-scale). After roughness analysis, samples were cleaned in ethanol once again using ultrasonic cleaner. Such prepared alloys were oxidized at 950 °C for 24 h in air using tube furnace. After exposure, samples were analyzed in terms of chemical composition using glow discharge optical emission spectrometry (GD-OES). The GD-OES depth profiles were quantified according to the procedure described in references [19–21]. After GD-OES analysis, samples were electro-plated with nickel and mounted in epoxy resin. Metallographic cross-sections of the oxidized alloy specimens were prepared by a series of grinding and polishing steps, the final step being fine polishing with the $SiO_2$ suspension with the 0.25 μm granulation. The cross-sections were analyzed using a Nikon Epiphot 300 optical microscope and a Hitachi S3400N scanning electron microscope (SEM). The oxidation kinetics was determined by oxide scale thickness measurement in 13 randomly selected locations on the cross-sections.

## 3. Results

### 3.1. Surface Roughness Evaluation

#### 3.1.1. Standard Roughness Evaluation

Figure 1 shows roughness profiles measured on the polished and ground surfaces of IN 625. One can note that the Y-axis scale is 10× higher for ground as compared with polished surface (Figure 1a,b respectively). The roughness profiles unambiguously show that polishing results in smoother surface, while grinding causes more rough surfaces. The roughness parameters determined based on standard roughness measurement were calculated based on Figure 1 and are shown in Table 2. The obtained roughness parameters, namely $R_a$, $R_z$, and $R_{max}$, clearly show that grinding increases values of mentioned parameters of about one order of magnitude in comparison to values obtained for polished surfaces. Moreover, there was a significant increase in roughness parameters, and the value of the standard deviation increases for ground as compared to polished surface.

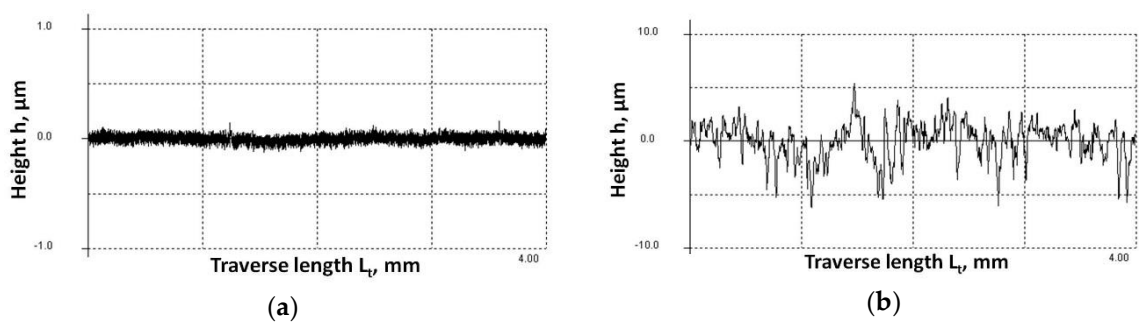

(a)          (b)

**Figure 1.** Surface roughness profiles performed by standard contact profilometer HOMMEL Werk T8000 on (**a**) polished (1 μm); (**b**) ground (220 grit) IN 625.

**Table 2.** Results of samples surface roughness evaluation using conventional procedures.

| Parameter | Polishing (1 μm) | | Grinding (220 grit) | |
|---|---|---|---|---|
| | Average Value | Standard Deviation | Average Value | Standard Deviation |
| $R_a$ | 0.039 | 0.007 | 1.067 | 0.065 |
| $R_z$ | 0.227 | 0.021 | 7.632 | 0.547 |
| $R_{max}$ | 0.288 | 0.056 | 9.189 | 0.800 |

3.1.2. Microroughness Evaluation Using Fractal Analysis

The plot showing relative length as a function of scale (expressed in μm) depicted in Figure 2 demonstrates the different shapes of the curves obtained for ground and polished surfaces of IN 625, namely, values of relative length for a given scale obtained for ground surfaces are higher as compared for the polished ones, especially for a smaller scale (see Figure 2). The parameters describing surface roughness based on fractal analysis, namely D, LSFC, and $L_R$, obtained for 1 μm scale are presented in Table 3 and show a similar trend as standard roughness parameters, i.e., they are higher for ground surface than the polished one. In contrast, the SRC value is smaller for the ground than polished surface.

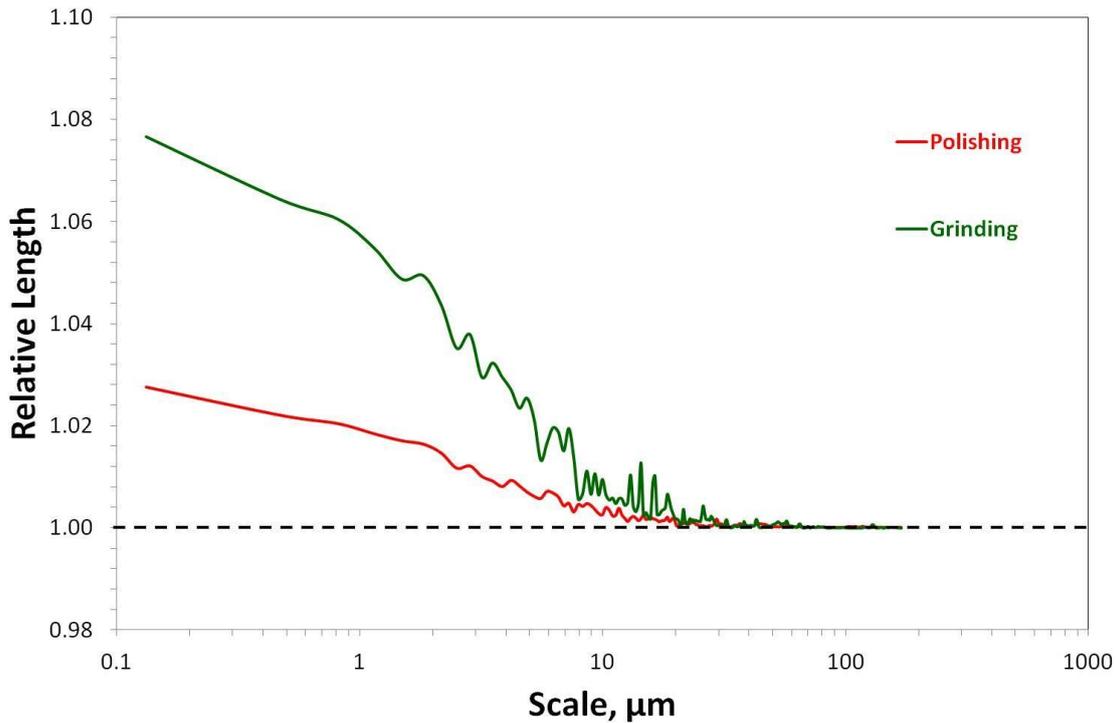

**Figure 2.** Relative length ($L_R$) as a function of scale plots obtained for studied surfaces by fractal analysis using Sfrax 1.0 software [16].

**Table 3.** Results of surface roughness evaluation using conventional procedures.

| Parameter | Polishing (1 μm) | Grinding (220 grit) |
|---|---|---|
| D | 1.0018 | 1.0084 |
| LSFC | 1.8 | 8.4 |
| $L_R$ 1 μm | 1.02 | 1.05 |
| SRC | 15.8 | 12.7 |

*3.2. Post-Exposure Analyses*

3.2.1. GD-OES Depth Profiles

Polished and ground IN 625 were air exposed at 950 °C for 24 h and analyzed using GD-OES. Figure 3 shows GD-OES profiles obtained on polished and ground IN 625. Profiles show the elemental concentration at given sputtering time, which in turn indirectly represents the distance from the surface (starting from 0 s which represents oxide scale surface). In general, one can observe that the sputtering time of the oxide scale formed on polished sample is higher that on ground (110 s and 70 s respectively). On both depth profiles a co-enrichment of Cr and O is observed during the first 100 s of measurement, which indicates that the majority of the oxide scales formed on both surfaces is $Cr_2O_3$. Moreover, on both profiles at the oxide scale/alloy interface (place where oxygen concentration drops) one can observe the enrichment of Nb. One can also observe a presence of the Ti in the outer oxide scale and below the oxide scale, Ti-depletion zone is observed. However, a difference in the depth profiles also exists between polished and ground material, namely, for the polished sample in the very outer part of the oxide scale (0–20 s), a strong enrichment of Mn (up to 10 at. %) and minor enrichment of Ni (up to 5 at. %) and Fe (up to 0.5 at. %) is observed. Conversely, on ground material, at the very outer part of the oxide scale, weaker enrichment of Mn is present (up to 5 at. %) and Ni is enriched up to about 0.9 at. %, while no enrichment of Fe is observed. Moreover, at the surface of the oxide scale together with Mn, enrichment of Ti is clearly visible (up to 1 at. %). In summary, in the case of a polished sample, a higher concentration of Mn and Ni is observed, while in the on ground sample, in the very outer part of the scale, a small peak from Ti is observed (up to 1 at. %). Moreover, plots show higher time of sputtering of the oxide scale on polished sample, which indicates that the oxide scale formed on polished surface is thicker than on the ground one.

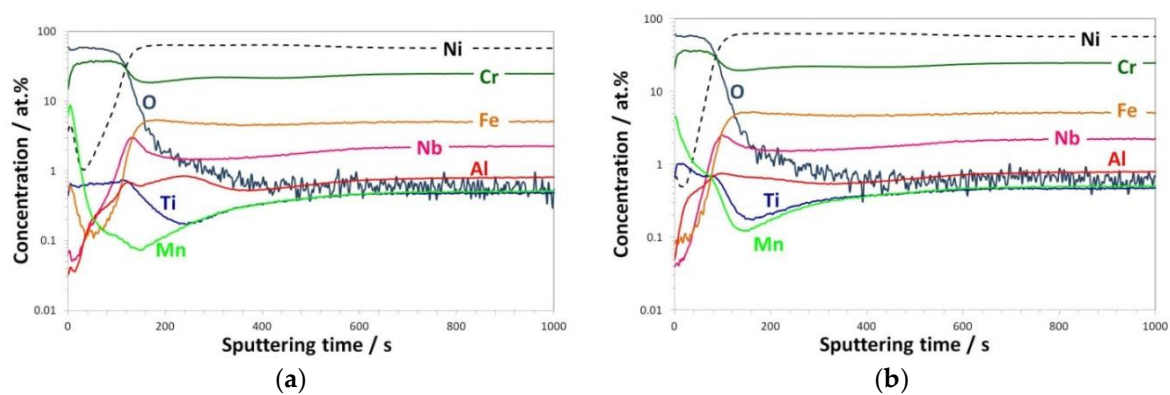

**Figure 3.** Quantified GD-OES depth profiles showing the atomic concentration of the given elements as a function of sputtering time determined for (**a**) polished (1 μm); (**b**) ground (220 grit) IN 625 after isothermal oxidation performed at 950 °C for 24 h in laboratory air.

3.2.2. SEM Analysis

The images of the cross-sections of polished and ground IN 625 after air exposure at 950 °C up to 24 h is presented in Figures 4a and 4b, respectively. From the images captured at lower magnification one can clearly see, that on the polished surface a thicker oxide scale was formed than on the ground sample. The SEM analysis performed at a higher magnification (inner images) showed, that on both samples, mainly $Cr_2O_3$ formed as an outer oxide scale, below which precipitates of internally oxidized aluminum are present. However, on top of the chromia scale formed on polished alloy, a layer of Cr, Mn, and Ni-mixed oxide is present. In both cross-sections, the Nb-rich phase, which most probably is $Ni_3Nb$, is present at the oxide scale/alloy interface, however this interface is more convoluted in the case of ground material (Figure 4b). Moreover, several voids are observed at the near-surface region on the ground sample. All observations using SEM are in a very good agreement with the results obtained using GD-OES. The average oxide scale thickness measured on the polished sample is 3.87 μm, while on ground it is 2.74 μm.

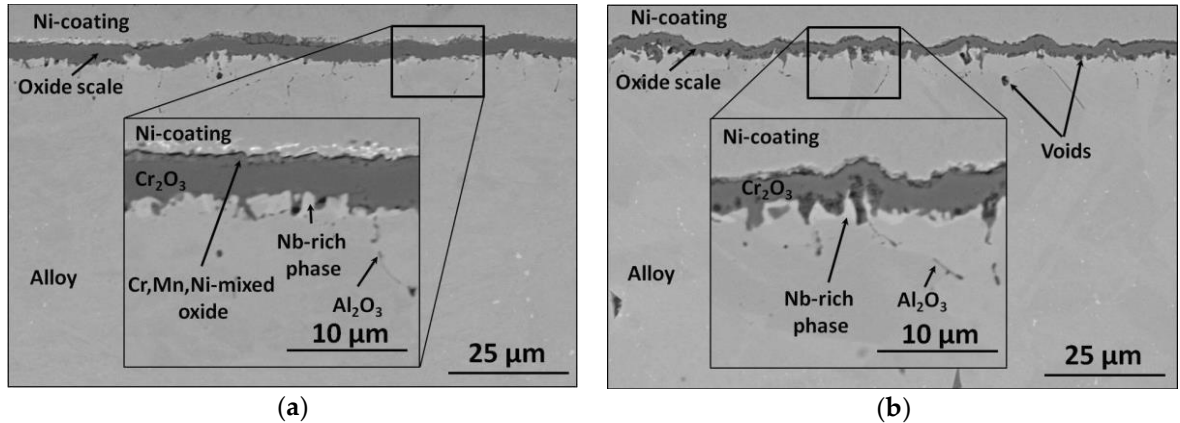

**Figure 4.** SEM/BSE images of (**a**) polished (1 μm); (**b**) ground (220 grit) IN 625 after isothermal oxidation performed at 950 °C for 24 h in laboratory air.

3.2.3. Oxidation Kinetics

The oxidation kinetics were determined using SEM images of the cross-sections of polished and ground alloy after exposure. The results are shown in Figure 5. The results show that the average value of the constant parabolic rate after 24 h of air exposure at 950 °C obtained for polished alloy is $2 \times 10^{-12}$ cm$^2$·s$^{-1}$, while for ground one it is $9 \times 10^{-13}$ cm$^2$·s$^{-1}$. The obtained results show that grinding decreased the oxidation kinetics of about twice.

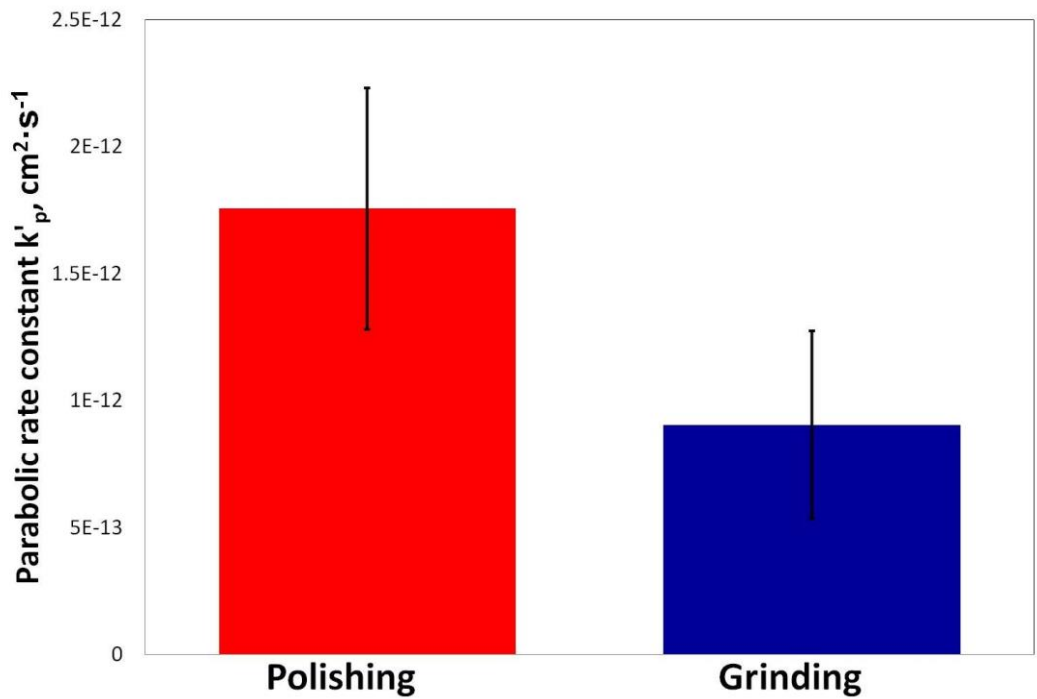

**Figure 5.** Parabolic rate constant k′$_p$ calculated based on the oxide scale thickness measurements on the studied alloy during isothermal exposure to laboratory air at 950 °C for 24 h.

**4. Discussion**

Both methods of roughness description, i.e., standard contact profilometry and fractal analysis, showed similar trends, specifically, roughness parameters where $R_a$, $R_z$, $R_{max}$, D, LSFC, and L$_R$ at 1 μm, obtained for ground surfaces, are substantially higher than for polished. In contrast, SRC values obtained for the ground surface are smaller than for the polished surface. All parameters clearly showed that surface preparation by grinding increased its roughness [15]. This in turn means, that the ground material possesses a higher surface to volume ratio than the polished. This observation is confirmed by the plot of relative length shown in Figure 2. Therefore, higher oxidation rate could be expected for ground material. Meanwhile, the results obtained in the present work showed exactly the opposite situation, where the material with the smaller surface to volume ratio showed a higher oxidation rate (Figure 5). The post-exposure analyses revealed the formation of Mn, Ni, and Fe-mixed oxide on top of the chromia scale (see Figures 3a and 4a). This mixed oxide originated most probably during the so-called "transient stage of oxidation", i.e., the very early stage of oxidation [22]. Contrary to the observation of polished material, on ground alloy formation of previously the mentioned Mn, Ni, and Fe-mixed oxide is not observed. This means, that the transient stage was much shorter in case of ground alloy. However, on top of the chromia scale developed on ground IN 625, slight enrichment of Ti is visible (Figure 3b). Ti is known to increase the chromia growth rate [23,24], however, in contrast to the material studied in references [23,24], in case of presently studied alloy, Ti content is only 0.4 wt %, and therefore, its influence on chromia growth rate is smaller. As described above, surface preparation changes the oxide scale composition and morphology, which in turn changes the oxidation kinetics at least during the early stage of high temperature oxidation of IN 625. The exact mechanism responsible for such change is not known so far. However, it was proposed by Nowak et al. [8] that during a different kind of surface preparation different amount/kind of defects are introduced into the near-surface region, namely, by the treatment resulting in higher surface roughness, a higher amount of defects are introduced. These defects are claimed to be easy diffusion paths of the elements forming protective oxide scales, as in the present case chromium. In Figure 4b, the presence of several voids was observed. In contrast, no such voids are observed in the case of the polished sample. The most probable formation of such voids is caused by the accumulation of introduced defects, as observed, for example, by Wierzba et al. [25]. However, to confirm the proposed mechanism, further investigations, including transmission electron microscopy (TEM) analysis and performing an oxidation tests using markers, are necessary and planned.

**Funding:** This research was financed within the Marie Curie COFUND scheme and POLONEZ program from the National Science Centre, Poland. POLONEZ grant No. 2015/19/P/ST8/03995. This project has received funding from the European Union's Horizon 2020 research and innovation programme under the Marie Skłodowska-Curie grant agreement No. 665778.

**Conflicts of Interest:** The author declares no conflict of interest. The funders had no role in the design of the study; in the collection, analyses, or interpretation of data; in the writing of the manuscript, and in the decision to publish the results.